\documentclass[%
 preprint,
superscriptaddress,
 amsmath,amssymb,
 aps,prl
]{revtex4-2}

\usepackage{xcolor}
\usepackage{graphicx}% Include figure files
\usepackage{dcolumn}% Align table columns on decimal point
\usepackage{bm}% bold math
\begin{document}

\preprint{APS/123-QED}

\title{Oxygen tilt-driven polar super-orders in BiFeO\textsubscript{3}-based superlattices}

\author{Ran Xu}
\email{ran.xu@centralesupelec.fr}
\author{Francesco Delodovici}
\author{Brahim Dkhil}
\affiliation{Universit\'{e} Paris-Saclay, CentraleSup\'{e}lec, CNRS, Laboratoire SPMS, 91190, Gif-sur-Yvette, France}
\author{Charles Paillard}
\email{paillard@uark.edu}
\affiliation{Department of Physics and Institute for Nanoscience and Engineering, University of Arkansas, Fayetteville, Arkansas 72701, USA}
\affiliation{Universit\'{e} Paris-Saclay, CentraleSup\'{e}lec, CNRS, Laboratoire SPMS, 91190, Gif-sur-Yvette, France}

%\affil[*]{ran.xu@centralesupelec.fr}
%\affil[**]{paillard@uark.edu}

%\keywords{Keyword1, Keyword2, Keyword3}

%\title{Manuscript Title:\\with Forced Linebreak}% Force line breaks with \\

%\affiliation{%


%\affiliation{
%\affiliation{
%\affiliation{%



\begin{abstract}
Ferroelectric-dielectric superlattices have attracted renewed interest for their ability to frustrate the polar order, leading to the emergence of exotic polar textures. The electrostatic depolarization, thought to be responsible for the complex polar textures in these superlattices can be alleviated by replacing the dielectric layer with a metallic one. One would thus expect that a close to uniform polarization state be recovered in the ferroelectric layer. However, here we show, using Density Functional Theory calculations, that {\color{blue} metastable} antipolar motions may still appear in superlattices combining multiferroic BiFeO\textsubscript{3} and metallic SrRuO\textsubscript{3} perovskite layers. We find that a complex oxygen octahedra tilt order, a so-called nanotwin phase, exists in BiFeO\textsubscript{3}/SrRuO\textsubscript{3} superlattices and competes with a more conventional phase. It leads to a doubling of the chemical period along the out-of-plane direction, owing to the presence of an oxygen octahedra tilt wave pattern and antipolar motions caused by trilinear energy couplings. We also show that out-of-plane polar displacements in the BiFeO\textsubscript{3} layer may reverse the (anti)polar displacements thanks to a strong quadrilinear coupling term. The oxygen tilt-driven couplings identified here {\color{blue} may open} new ways to engineer and control polar displacements in superlattice based polar metals and hybrid improper (anti)ferroelectrics.

%\begin{description}
%\end{description}
\end{abstract}

\maketitle

%\tableofcontents

%\section{\label{sec:level1} Introduction}

%{\color{blue}
%Metastable states with low lying energy 
%}
Ferroelectric superlattices (SLs) are repeated stacking of alternating ferroelectric nanolayers and dielectric or metallic layers. SL architectures allow to control both  mechanical and electrical boundary conditions felt by the ferroelectric nanolayers. 
%Deposition techniques with atomic layer control have allowed to explore the rich playground that ferroelectric and multiferroic SLs represent. 
Intriguing new physics has resulted from {\color{blue} exploring ferroelectric SLs and nanostructures, such as } the presence of polar, topologically protected quasi-particles~ \cite{yadav2016observation,das2019observation,hong2017stability,prosandeev2008control,nahas2015discovery,choudhury2011geometric,prosandeev2008controlling}, {\color{blue} often emerging as low energy metastable states~\cite{nahas2015discovery,Prokhorenko2017,Goncalves2019}. In fact, metastable phases have been wildly evidenced in ferroelectric nanostructures and manipulated, for instance with strain, electric fields~\cite{caretta2023non,Li2017} or optical excitation~\cite{stoica2019optical,dansou2022controlling} to achieve new exotic properties such as negative capacitance~\cite{zubko2016negative,das2021local,walter2020strain} when reaching these hidden phases.} 

%Other works have evidenced the emergence of negative capacitance effects in the frustrated ferroelectric layers of PbTiO\textsubscript{3}/SrTiO\textsubscript{3} or BaTiO\textsubscript{3}/SrTiO\textsubscript{3} superlattices\cite{zubko2016negative,das2021local,walter2020strain}. At last, optical stimulation of these SLs has revealed new hidden phases, including a so-called supercrystal phase \cite{stoica2019optical,dansou2022controlling}.
In most {\color{blue}ferroelectric/dielectric SLs}, a uniformly out-of-plane polarized ferroelectric nanolayer would experience a large depolarizing electric field, resulting from poor electrostatic screening of the polarization bound charges by the dielectric layer. Electrostatic frustration was thus put forward as an explanation of the resulting structure, and thus functional properties \cite{hong2017stability,hong2018blowing,bennett2020electrostatics} of ferroelectric SLs. It is however legitimate to ask whether {\color{blue} other degrees of freedom, such as oxygen octahedra tilts in perovskite oxides, play an important role in the formation of complex structural phases in SLs. One way to test this hypothesis is to employ a \textit{metallic} spacing layer rather than a dielectric one, thus limiting or cancelling electrostatic depolarizing effects.

In this regard, we mostly focus on BiFeO\textsubscript{3}/SrRuO\textsubscript{3} SLs.} BiFeO\textsubscript{3} (BFO) is a prototypical multiferroic with large spontaneous polarization,  antiferromagnetic order superimposed with a cycloidal spin modulation and strong $a^-a^-a^-$ oxygen octahedra tilt pattern in Glazer notation~\cite{glazer1972classification} at room temperature \cite{lebeugle2007very,lebeugle2007room,ramazanoglu2011local,sosnowska1982spiral}.
Recent works have reported emerging complex phases in BFO-based superlattices, such as antiferroelectric phases in BiFeO\textsubscript{3}/LaFeO\textsubscript{3}~\cite{carcan2018interlayer} and BiFeO\textsubscript{3}/NdFeO\textsubscript{3}~\cite{khaled2021anti} SLs. Perhaps most strikingly, BiFeO\textsubscript{3}/TbScO\textsubscript{3} SLs have shown {\color{blue} the room temperature coexistence of a complex polar phase and an antiferroelectric \textit{Pnma} phase as well as their electrical control~\cite{caretta2023non}.} In addition, BFO has a rich polymorphic playground, with for instance a low energy lying \textit{Pnma} phase \cite{dieguez2011first,Mundy2022Liberating} {\color{blue} with $a^{-}a^{-}c^{+}$ tilts}. It can thus be hoped that a large number of phases, and properties, can be addressed even in BFO/metal SLs {\color{blue} if one manages to frustrate its octahedra rotation pattern, for instance by associating BFO with a perovskite metal having an $a^-a^-c^+$ tilt pattern such as SrRuO\textsubscript{3} (SRO)~\cite{bansal2003metal,randall1959preparation}.}  

The present work uses Density Functional Theory (DFT) calculations {\color{blue} to explore the impact of oxygen octahedra tilt rotation on the emergence of complex phases in BFO-based SLs}. We show that, despite the metallic nature of SRO, {\color{blue} which limits depolarizing effects,} unexpected in-plane antipolar motions may be retained in the SL in a competing super-ordered phase. We attribute this result to the strong trilinear coupling between Bi cations motion and oxygen octahedra tilts. Concurrently, polar displacements in the out-of-plane direction are retained in the BFO layer due to screening of the polarization charges by the metallic SRO layer. We show that this out-of-plane polar displacement may help control the direction of (anti)polar in-plane atomic motions thanks to quadrilinear coupling involving oxygen octahedra tilts.

%\section{\label{sec2} Methods}

DFT calculations {\color{blue}were performed using} the Vienna Ab-initio Simulation Package \cite{kresse1993ab,kresse1994ab,kresse1996efficiency,blochl1994projector} with the Projector Augmented Waves method \cite{blochl1994projector,kresse1999ultrasoft}. Our pseudo-potentials include valence electrons from Bi $5d$, $6s$ and $6p$, Fe $3s$, $3p$, $3d$ and $4s$, Sr $4s$, $4p$, $5s$, Ru $4s$, $4p$, $4d$, $5s$ and O $2s$, $2p$ states. We employ the PBESol exchange-correlation functional~\cite{perdew2008restoring}. Following the literature, we apply a Hubbard correction~\cite{dudarev1998electron} of 4 eV and 0.6 eV on the $d$ orbitals of Fe~\cite{dieguez2011first} and Ru~\cite{rondinelli2008electronic,menescardi2021comparative} atoms. 
%, as formulated by Dudarev et al.\cite{dudarev1998electron}. 
Collinear magnetism is assumed. The plane wave cut-off is 500 eV, meanwhile {\color{blue} a $5\times 5 \times 5$ Monkhorst-Pack mesh~\cite{monkhorst1976special} is employed.}
%the Brillouin zone is sampled using a $5\times 5 \times 5$ Monkhorst-Pack mesh~\cite{monkhorst1976special}. 
{\color{blue}Total energy is converged below $10^{-7}$~eV in self-consistent cycles.}
%We consider self consistency achieved when changes in the total energy are smaller than 10$^{-7}$ eV. 
Structural convergence is achieved when the forces 
%acting on the ions 
are smaller than 2 meV/\AA. %Post-processing of DFT data was performed via Vaspkit \cite{wang2021vaspkit}.
{\color{blue} To mimic the effect of epitaxial strain imposed by a cubic SrTiO\textsubscript{3} substrate, we fix the in-plane lattice constants of the SLs to 3.895 \AA, as calculated from DFT.}
%We fixed the in-plane lattice constants to 3.895 \AA (), the calculated relaxed lattice constant of cubic SrTiO\textsubscript{3}, to mimic the effect of epitaxial strain imposed by a cubic SrTiO\textsubscript{3} substrate. 
{\color{blue} At the SrTiO\textsubscript{3} in-plane lattice constant (and in general reasonable strains in the range -2\% to +2\%), we do not expect strained BiFeO\textsubscript{3} and SrRuO\textsubscript{3} to exhibit markedly different structural properties from the bulk according to the literature~\cite{Sando2016, Zayak2008}.}
Phonon band structures for high-symmetry (cubic-like) BFO/SRO SLs were obtained using density functional perturbation theory as implemented in VASP and the Phonopy package~\cite{togo2015first}.

%\section{\label{sec3} Results }

\begin{figure}[]
\centering
\includegraphics[width=0.95\linewidth]{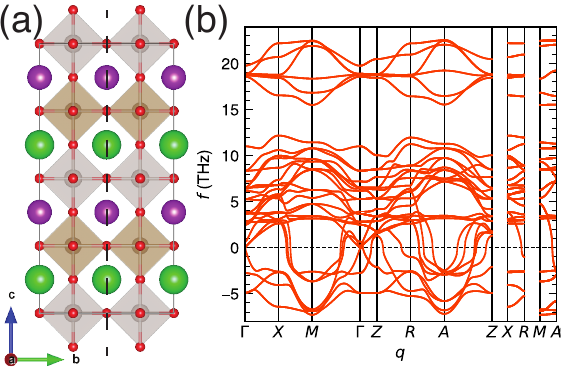}
\caption{Phonon instabilities in high-symmetry superlattices [BiFeO\textsubscript{3}]\textsubscript{1}/[SrRuO\textsubscript{3}]\textsubscript{1} superlattices (a) Sketch of the $2\times2\times2$ supercell with atoms in their cubic-like, high symmetry positions. (b) Phonon dispersion in the cubic-like phase (high-symmetry points in the first Brillouin zone are defined in the Supplemental Material~\cite{SMxu2024}).}
\label{fig:phonons}
\end{figure}

\begin{figure}[]
\centering
\includegraphics[width=0.75\linewidth]{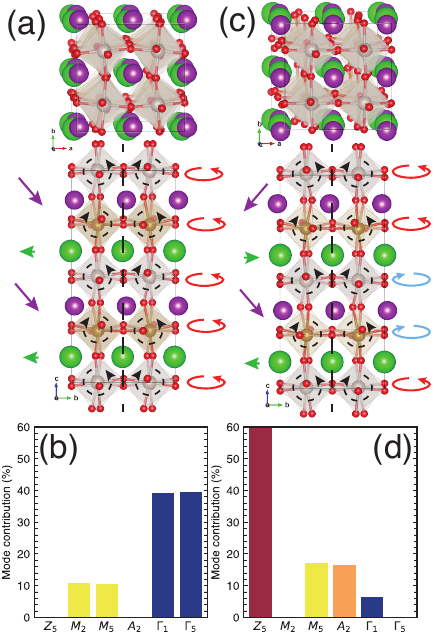}
\caption{Atomic patterns in conventional and super-ordered phases (a) Ground state relaxed structure and (b) its projection on phonon modes of the high-symmetry structure; (c) super-ordered relaxed structure and (d) its projection on phonon modes of the high symmetry structure. {\color{blue} Arrows on the left of each structure depict A-site cation displacements, while rotating arrows on the structure depict $M_5$-related oxygen tilts. Right side rotating arrows depict clockwise (blue) or anti-clockwise oxygen octahedra rotation associated with $M_2$ (a) and $A_2$ (c) modes.}}
\label{fig:projections}
\end{figure}

\begin{figure}[]
\centering
\includegraphics[width=0.75\linewidth]{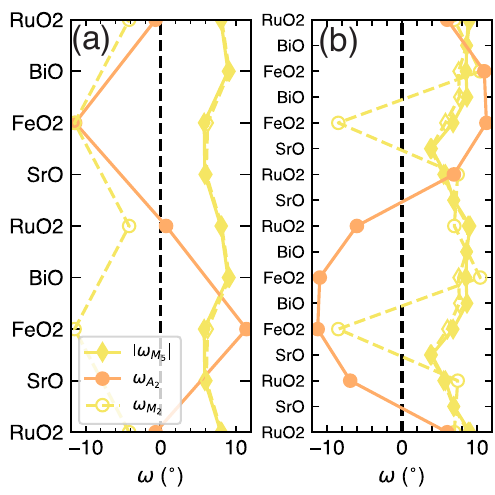}
\caption{Complex tilt patterns in BFO/SRO superlattices. (a) BFO\textsubscript{1}/SRO\textsubscript{1} in-plane tilt angle (yellow diamonds) associated with the $M_5$ mode and out-of-plane tilts (circles) associated with the $M_2$ (yellow) and $A_2$ (orange) modes. The conventional and super-ordered phase are depicted by dashed and plain lines respectively. (b) Oxygen octahedra tilt angles in the conventional (dashed lines) and super-ordered (plain lines) phase of BFO\textsubscript{2}/SRO\textsubscript{2}, with the same color code as (a).}
\label{fig:tilts}
\end{figure}

%\subsection*{High symmetry phase and instabilities}
We start by calculating the phonon dispersion of [BiFeO$_3$]$_1$/[SrRuO$_3$]$_1$ SL (subsequently noted BFO\textsubscript{1}/SRO\textsubscript{1}), with all ions fixed in the high-symmetry {\color{blue}positions} of the cubic perovskite structure (see Figure~\ref{fig:phonons}a). 
The high-symmetry structure has \textit{P4mm} space group due to the chemical arrangement. The phonon dispersion, shown in Figure~\ref{fig:phonons}b, shows strong imaginary frequencies (depicted as negative), a hallmark of major structural instabilities. The most prominent instabilities exist at the $M$ $(1/2,1/2,0)$ and $A$ $(1/2,1/2,1/2)$ points in the Brillouin zone. The lowest-frequency unstable $M$ mode corresponds to antiphase oxygen octahedra tilts along the out-of-plane direction (see Supplemental Material~\cite{SMxu2024}). We adopt a Glazer-like notation $a^{-/+}b^{-/+}c^{-/+}$ for each pseudo-cubic perovskite cell, where -/+ indicates anticlockwise and clockwise rotation of the oxygen octahedron. We can then describe the lowest unstable $M$ mode by the sequence $00c^{+}$/$00c^{\prime +}$  for the BFO\textsubscript{1}/SRO\textsubscript{1} SL. The unstable mode at the $A$ point, which has $A_2$ symmetry~\cite{SMxu2024}, corresponds to a complex wave-like arrangement of oxygen octahedra tilts, leading to  doubling of the SL period (see Figure~\ref{fig:projections}c and Figure~\ref{fig:tilts}). In our Glazer-like notation, this mode corresponds to a  tilt pattern $00c^{-}/00c^{\prime -}/00c^{+}/00c^{\prime +}$. Concurrently, less unstable modes at $\Gamma$ (see Figure~\ref{fig:phonons}b) show polar displacements carried by off-centering of the Bi ions in-plane ($\Gamma_5$ symmetry) and out-of-plane ($\Gamma_1$ symmetry) respectively. Interestingly, the $\Gamma-Z$ lowest unstable branch is flat, indicating that polar motions of the Bi ions between two BFO layers separated by a SRO layer do not carry an additional electrostatic cost compared to their antipolar displacement. The metallic nature of the SRO therefore likely screens the electrostatic dipolar energy cost associated with in-plane polar motions, and thus effectively decouples the in-plane polar motions of successive BFO layers. Note that the $\Gamma-Z$ branch flatness also suggests the possibility to access polar states which combine multiple k-points, as can be the case for polar skyrmions or vortices~\cite{yadav2016observation,das2019observation}.
%

%\subsection*{Relaxed structures}

Next, we relax the full BFO\textsubscript{1}/SRO\textsubscript{1} superlattice in a $2\times2\times4$ pseudocubic supercell (effectively simulating a BFO\textsubscript{1}/SRO\textsubscript{1}/BFO\textsubscript{1}/SRO\textsubscript{1} arrangement in the out-of-plane direction), {\color{blue} thus allowing} the significant $A$ and $Z$ instabilities to develop and double the chemical wavelength. {\color{blue} After} exploring various atomic distortions starting points, we eventually find two structures with minimal energy. The first one, the ground state, is depicted in Figure~\ref{fig:projections}a. 
It is characterized by (1) an $a^-a^-c^+$ general tilt system in Glazer notation and (2) out-of-plane and in-plane polar motions in the BFO layer along the $[00\bar{1}]$ and $[110]$ directions respectively. {\color{blue} Out-of-plane polar motions show displacements of the Bi ions towards FeO\textsubscript{2} planes, and have $\Gamma_1$ symmetry. They represent $\approx40$\% of the distortions (Figure~\ref{fig:projections}b). In-plane polar motions, of $\Gamma_5$ symmetry, consist mostly of opposite motions of Bi and O along the $[110]$ direction and represent 40\% of the distortions. Small Sr and O motions, opposite to the BiO plane motions, are also present, }a feature reminiscent of hybrid improper ferroelectrics~\cite{rondinelli2012octahedral,Mulder2013Turning}, and whose origin can be traced back to known atomistic couplings between dipolar displacements and oxygen octahedra tilts in perovskite oxides \cite{Bellaiche2013Universal}. Of course, the present SLs are metallic in the RuO\textsubscript{2} planes (see Supplemental Material~\cite{SMxu2024}). Thus the {\color{blue} the ground state (which we coin conventional phase)} is not ferroelectric but exhibits polar features. Octahedra tilts account for the remaining 20\% of the structural distortions, with $M_5$ ({\color{blue} $a^-a^-0$} tilts of amplitude $6-8^{\circ}$, see Figure~\ref{fig:tilts}a) and $M_2$ ({\color{blue}$00c^+$ tilts with typical amplitude of $11.5^{\circ}$ in the BFO layer and $4^{\circ}$ in the SRO layer}) modes each contributing to about 10\% of the distortion. In comparison, our calculated value for bulk octahedra rotation in BFO and SRO are respectively $12.6^{\circ}$ and $7^{\circ}$. %The $M_2$ tilts along the out-of-plane direction have a . 
%

%As a result, the $M_2$ mode can be thought of as a linear combination of in-phase $00c^+$  tilts with amplitude $7.75^{\circ}$ and out-of-phase $00c^-$  tilts with amplitude $3.75^{\circ}$ (see Figure~\ref{fig:tilts}).
%

%We project the distortions of the calculated ground state onto the phonon modes of the high symmetry structure~\cite{SMxu2024}. The polar distortions, as shown in Figure~\ref{fig:projections}b, account for approximately 80\% of the structural distortions.
%
%
%The $\Gamma_1$ (out-of-plane polar motion) and $\Gamma_5$ (in-plane polar motion) have approximately equal weight. The $\Gamma_1$ mode distortion is dominated by ionic motions of the Bi ions towards the FeO$_2$ plane. Meanwhile the $\Gamma_5$ mode consists of opposite in-plane motion of Bi and O along the $[110]$ direction, as well as Sr and O motion in the SrO plane (moving opposite to Bi and O in BiO planes), a feature reminiscent of hybrid improper ferroelectrics~\cite{rondinelli2012octahedral,Mulder2013Turning}, and whose origin can be traced back to known atomistic couplings between dipolar displacements and oxygen octahedra tilts in perovskite oxides \cite{Bellaiche2013Universal}. Octahedra tilts account for the remaining 20\% of the structural distortion, with $M_5$ mode (anti-phase in-plane tilts of about $6-8^{\circ}$, see Figure~\ref{fig:tilts}) and $M_2$ each contributing to about 10\% of the distortion. The $M_2$ tilts along the out-of-plane direction have a typical amplitude of $11.5^{\circ}$ in the BFO layer and $4^{\circ}$ in the SRO layer. As a result, the $M_2$ mode can be thought of as a linear combination of in-phase $00c^+$  tilts with amplitude $7.75^{\circ}$ and out-of-phase $00c^-$  tilts with amplitude $3.75^{\circ}$ (see Figure~\ref{fig:tilts}).
%
%To compare with, our calculated value for bulk octahedra rotation in BFO and SRO are respectively $12.6^{\circ}$ and $6.9-7^{\circ}$.
%

Surprisingly, our relaxation evidenced a second structure with very close energy to the ground state (9 meV/perovskite cell), depicted in Figure~\ref{fig:projections}c. We observe that the structural period is doubled out-of-plane with respect to the superlattice chemical period. We refer to this structure as “super-ordered”. The super-ordered structure {\color{blue} has similar $M_5$-symmetry $a^-a^-0$ tilt} pattern and amplitude as the ground state structure (see Figure~\ref{fig:tilts}a). It possesses, as well, {\color{blue} similar} out-of-plane polar motions of the Bi ions, albeit with smaller amplitude than the conventional phase. {\color{blue}The main differences between the conventional and super-ordered structure} arise from the condensation of a tilt wave-like pattern of $A_2$ symmetry (see Figure~\ref{fig:projections}c and Figure~\ref{fig:tilts}) and in-plane antipolar displacements of $Z_5$ symmetry. {\color{blue} The $Z_5$ displacements resemble the $\Gamma_5$ displacements of the conventional phase, but reverse sign every SL chemical period (see Figure~\ref{fig:projections}c). They account for 60\% of the super-ordered structure distortions. The $A_2$ mode, a tilt wave-like pattern where rotations around the out-of-plane axis alternate between clockwise and anticlockwise every period (see Figures~\ref{fig:projections}c\&\ref{fig:tilts}a), represents about 14\% of the total distortions (Figure~\ref{fig:projections}d).} It is an instance of the nanotwin phases predicted to occur in BFO at high temperature~\cite{prosandeev2013Novel} or in BiFeO\textsubscript{3}/NdFeO\textsubscript{3} solid solutions \cite{Xu2015Finite}, and generates the antipolar displacement pattern $Z_5$ arising from trilinear couplings in the free energy landscape of the form $M_5 A_2 Z_5$ (see below, and Ref.~\cite{Bellaiche2013Universal}).

{\color{blue} Whether combining perovskites with competing tilt systems in superlattices universally leads to the existence of (meta)stable super-ordered phases likely relies on the relative strength of the tilt instabilities in the high-symmetry phase of the perovskites composing each nanolayer. Yet, our work shows that BFO-based SLs are an interesting playground to engineer such super-ordered phases, as recent experimental reports have found some evidence of their existence via High-Resolution Transmission Electron Microscopy~\cite{gu2024} or X-Ray diffraction~\cite{Maran2014,Maran2016}. In addition, we predict that (1) larger SLs, such as BFO$_2$/SRO$_2$, also exhibit metastable super-order tilt-wave-like patterns associated with antipolar features (see Figure~\ref{fig:tilts}b and Supplementary Material~\cite{SMxu2024}); (2) BFO/dielectric SLs, such as BiFeO\textsubscript{3}/LaFeO\textsubscript{3} (BFO/LFO), also harbor such metastable super-ordered phases (see Supplemental Material~\cite{SMxu2024}), consistent with recent observations~\cite{gu2024} indicating competing conventional and super-ordered phases in BFO\textsubscript{n}/LFO\textsubscript{n} up to $n=5$.}

%such as BiFeO\textsubscript{3} ($a^{-}a^{-}a^{-}$) }
%Our work thus indicates that these super-orders may be universal features of BFO-based superlattices when the BFO and dielectric or metallic layer have competing tilt systems.

%\section{\label{sec4} Discussion}

Since the “conventional” (Figure~\ref{fig:projections}a) and “super-ordered” (Figure~\ref{fig:projections}c) phases {\color{blue} are so close in energy (9~meV/p.u.)}, it is {\color{blue} possible} that they coexist at room temperature {\color{blue}($\frac{k_B T}{2} \approx 12.5$~meV at room temperature), as coexisting metastable phases have already been observed in some BFO-based superlattices~\cite{caretta2023non,gu2024}}. {\color{blue} It is also known that external stimuli such as THz or visible excitation~\cite{Nova2019}, thermal quenching~\cite{Nahas2020} or electric field application~\cite{caretta2023non} (for dielectric superlattices such as BFO/LFO) may be employed to access close in energy metastable states}. We now set out to understand the atomistic energy landscape explaining the emergence of this super-ordered phase in BFO$_1$/SRO$_1$ {\color{blue} using symmetry-relevant modes and a Taylor expansion of the energy around the high-symmetry structure of Figure~\ref{fig:phonons}a. Such approach has been successfully applied over the years to derive effective Hamiltonians in complex perovskite oxides such as BiFeO\textsubscript{3}~\cite{Kornev2007} and related superlattices~\cite{Zanolli2013} providing excellent agreement with experimental observations~\cite{Infante2010,Rispens2014}}.
Based on the projection of the atomic displacements onto the phonon calculated in Figure~\ref{fig:phonons}, we were able to construct a set of six symmetry-adapted characteristic displacement patterns: $u_{\Gamma_1}$ represents polar displacements of the Bi sublattice in the [001] direction; $u_{\Gamma_5}$ represent antipolar motions of Bi and Sr ions along the [110] direction and it is akin to polar displacements found in hybrid improper ferroelectrics~\cite{rondinelli2012octahedral}. Meanwhile $u_{Z_5}$ indicates that these latter displacement change sign every chemical period along the out-of-plane direction; $\phi_{M_2}$ and $\phi_{A_2}$ represent the oxygen tilting pattern around $[001]$ summarized in Figure~\ref{fig:tilts}; $\phi_{M_5}$ represents an $a^-a^-0$ pattern of oxygen octahedra rotations.
Both the conventional and super-ordered phases show the $M_5$ tilts with similar magnitude (see Figure~\ref{fig:tilts}), as well as the $\Gamma_1$ polar out-of-plane polar motion of the Bi ions towards the FeO\textsubscript{2} plane. 
%Condensing those 2 modes from the high symmetry \textit{P4mm} structure shows that they lower energy the most (see Figure~\ref{fig:energetics}). Interestingly enough, $u_{\Gamma_1}$ only favors displacements of Bi towards the FeO\textsubscript{2} plane. 
%
\begin{figure}[]
\centering
\includegraphics[width=0.85\linewidth]{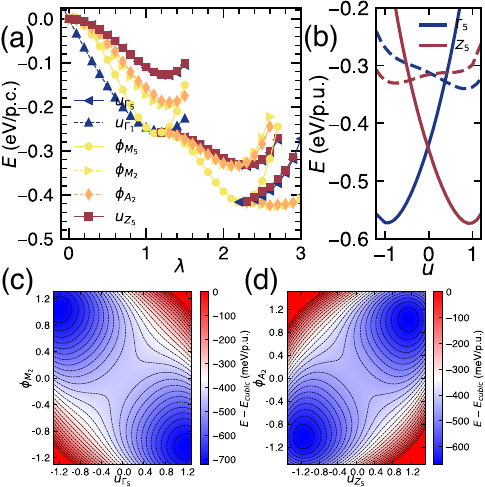}
\caption{Energetics and potential for polar displacement switching (a) Energy with respect to the high symmetry phase, when modes are condensed. (b) Energy with respect to the high symmetry phase when $u_{Z_{5}}$ (red) is condensed while $\phi_{M_{5}},\phi_{A_{2}} = 1$ and when $\Gamma_5$ (blue) is condensed while $\phi_{M_{2}},\phi_{M_{5}} = 1$. Dashed lines indicate no out-of-plane polar displacement ($u_{\Gamma_{1}} = 0$), while plain lines indicate $u_{\Gamma_{1}} = 1$; (c) and (d) are ($\phi_{M_{2}},\phi_{\Gamma_{5}})$ and ($\phi_{A_{2}},\phi_{Z_{5}}$) energy maps when $u_{\Gamma_1} = 1$ and $\phi_{M_5} = 1$ .}
\label{fig:energetics}
\end{figure}

To further elucidate the origin of the structural features of BFO/SRO SLs, we condense individually the displacement patterns of different symmetry to understand the energy couplings at play. Clearly, the $\Gamma_1$ out-of-plane polar mode and $M_5$ in-plane tilt modes are the strongest instabilities, leading each to a lowering of the energy from the high symmetry structure by about 250 meV/p.u. each (see Figure~\ref{fig:energetics}a). Subsequently, once the $\Gamma_1$ and $M_5$ modes are condensed, only the oxygen octahedra rotation along the out-of-plane direction lower the energy, with the $M_2$ and $A_2$ modes slightly lowering the energy further. Only then, once the modes $\Gamma_1$,$M_5$ and either $M_2$ or $A_2$ are condensed, can the energy be further lowered by condensing the in-plane polar $\Gamma_5$ mode (when $M_2$ is condensed) or the in-plane antipolar mode $Z_5$ (when $A_2$ is condensed) as plotted in Figure~\ref{fig:energetics}b). Interestingly, Figure~\ref{fig:energetics}b-d shows that the energy curve is asymmetric; symmetry analysis of possible energy couplings (see Supplemental Material~\cite{SMxu2024}) reveals that it is the result of trilinear coupling terms of the form $\phi_{M_2}\phi_{M_5}u_{\Gamma_5}$ and $\phi_{A_2}\phi_{M_5}u_{Z_5}$. Quite surprisingly, the out-of-plane polar mode $\Gamma_1$ significantly alters these trilinear couplings: without it ($u_{\Gamma_1}=0$, see dashed lines in Figure~\ref{fig:energetics}b) the minimum of energy for the $Z_5$ or $\Gamma_5$ mode is reversed compared to the case $u_{\Gamma_1}=1$. This quadrilinear coupling with out-of-plane polar displacements creates new opportunities to use BFO-based SLs to engineer switchable hybrid improper polar displacements. Polarization is notoriously difficult to switch in hybrid improper ferroelectrics. Since we have indications that the features exposed in this work apply to other BFO-based SLs~\cite{gu2024}, one may envision new pathways to switch polarization in hybrid improper ferroelectrics. In particular, in BFO/dielectric superlattices, one could imagine to manipulate $u_{\Gamma_1}$ via electric fields or optical excitation, and leverage the mechanisms demonstrated in Figure~\ref{fig:energetics}b to eventually switch the in-plane polarization. In fact, we show in the Supplementary Material that $u_{\Gamma_1}$ can indeed control the direction of in-plane polar moments in [BiFeO\textsubscript{3}]\textsubscript{1}/[LaFeO\textsubscript{3}]\textsubscript{1} SLs.

%\section{Conclusion}
The present work, by means of ab-initio calculations, reveals that tilt-induced nanotwin super-orders can be engineered in multiferroic BiFeO\textsubscript{3}-based SLs. We also show that, even in BFO/metal SLs, one may generate antipolar displacements due to trilinear couplings acting in the nanotwin phase. Furthermore, the coupling between the out-of-plane polar displacements, in-plane (anti)polar and tilts {\color{blue} may be a step towards} control of the polarization in hybrid improper ferroelectrics of hybrid improper polar metals, for instance via electrical or optical means,{\color{blue} or THz manipulation using the squeezing effect to reduce the out-of-plane polarization~\cite{Chen2022}. Future works will investigate how to further stabilize these super-ordered phases and manipulate the polar order in various BFO-based superlattices. A promising prospect is the use of tensile bi-axial strain, as shown in the Supplemental Material~\citep{SMxu2024}.}

\section*{Acknowledgements}
R. X. and C. P. thank the China Scholarship Council (CSC). C.P. acknowledges financial support from Agence Nationale de la Recherche (ANR) under Grant No. ANR-21-CE24-0032 (SUPERSPIN). We thank the European Union’s Horizon 2020 research and innovation program under Grant Agreement No. 964931 (TSAR). This work was performed thanks to the Mesocentre Ruche computing center from Université Paris-Saclay and the TGCC high-performance computing center through grant A0130912877. We thank Pr. L. Bellaiche for insightful comments on this manuscript.

%\section*{Author contributions statement}

%R.X., F.D. and C.P. performed simulations. C.P. conceived and supervised the study. R.X. and C.P. wrote the first version of the manuscript. All authors analyzed and discussed the results, reviewed and revised the manuscript. 

%\section*{Additional information}
%The authors declare no competing interests.

\appendix

%\section{Methods}
%\section*{Methods}

\bibliography{sample}% Produces the bibliography via BibTeX.

\end{document}